\documentclass[aps,showpacs,nofootinbib]{revtex4}

\usepackage{graphicx}
\usepackage{graphics}
\usepackage{amssymb}
\usepackage{amsmath}
\usepackage{bm}
\newcommand{\be}{\begin{equation}}
\newcommand{\ee}{\end{equation}}
\newcommand{\bea}{\begin{eqnarray}}
\newcommand{\eea}{\end{eqnarray}}
\newcommand{\beal}{\begin{align}}
\newcommand{\eal}{\end{align}}
\newcommand{\bespl}{\begin{split}}
\newcommand{\espl}{\end{split}}

\newcommand{\nslash}{\kern 0.2 em n\kern -0.50em /}
\newcommand{\kslash}{\kern 0.2 em k\kern -0.45em /}
\newcommand{\pslash}{\kern 0.2 em p\kern -0.50em /}
\newcommand{\Sslash}{\kern 0.2 em S\kern -0.50em /}
\newcommand{\Pslash}{\kern 0.2 em P\kern -0.50em /}
\newcommand{\Rslash}{\kern 0.2 em R\kern -0.50em /}
\begin{document}

\title{
Partial restoration of factorization and universality 
in presence 
of factorization breaking interactions in 
hadronic hard scattering processes.  
}

\author{A.~Bianconi}
\email{andrea.bianconi@bs.infn.it}
\affiliation{Dipartimento di Chimica e Fisica per l'Ingegneria e per i 
Materiali, Universit\`a di Brescia, I-25123 Brescia, Italy, and\\
Istituto Nazionale di Fisica Nucleare, Sezione di Pavia, I-27100 Pavia, Italy}

%%%%%%%%%%%%%%%%%%%%%%%%%%%%%%%%%%%%%%%%%%%%%%%%%%%%%%%%%%%%
\begin{abstract}
Recent works have discussed the violation of factorization and universality 
in hadronic hard scattering processes aimed at measurements of 
T-odd distributions. We use simple arguments  
to show that it is possible 
to restore an approximate factorization involving T-odd 
contributions, if the factorization 
breaking interactions present a frequency spectrum 
dominated by a narrow and regular peak whose maximum value  
corresponds to a respected factorization. 
\end{abstract} 


\maketitle

\section{Introduction}

Recent works\cite{BBMP,CollinsQiu07} 
have seriously re-discussed the ideas of factorization 
and universality in hard processes aimed at  
the detection of T-odd distributions. 
In the last fifteen years the study 
of these distributions 
has often made it necessary to reconsider the basic properties 
of the QCD-improved parton 
model, in particular 
factorization\cite{CollinsSoperSterman,Bodwin}. 

In quark/parton physics the first T-odd 
distribution function was probably 
the Sivers function\cite{Sivers} in 1990, followed by the 
Boer-Mulders-Tangerman 
function\cite{MuldersTangerman96,BoerMulders98,Boer99}. 
The former was used to explain single spin 
asymmetries\cite{ABM95}, the latter to explain 
unpolarized Drell-Yan azimuthal 
asymmetries\cite{Boer99}. 
Recently, it has been demonstrated\cite{JQVY06} that the T-odd 
mechanism introduced in\cite{EfremovTeryaev82} and \cite{QiuSterman91} 
produces a Sivers asymmetry when its effect is extrapolated into the 
small transverse momentum region. 
Also, in high energy nuclear physics a T-odd 
structure function, the so-called ``fifth structure function'',  
was introduced\cite{Donnelly} 
and modeled\cite{BB95,BR97} to describe normal asymmetries 
in $A(\vec e,\vec e' p)$ quasi-elastic scattering.  

For justifying the same existence of leading twist 
T-odd functions, the central object is the gauge restoring 
operator entering the definition of parton 
distribution\cite{CollinsSoperSterman}. In an approach 
where factorization is assumed from the very beginning, this gauge 
field is assumed to be negligible at leading twist, but in its 
absence T-odd distributions  
are  forbidden\cite{Collins93} by general 
invariance principles. 
A chain of arguments and 
examples\cite{BrodskyEtAl02,BrodskyHwangSchmidt02,Collins02,JiYuan02, 
BoerBrodskyHwang03,BelitskyJiYuan03} 
has shown that 
proper taking into account the gauge factor permits the existence of 
leading  
twist T-odd distributions in a QCD framework, associated with 
formal factorization breaking by interactions in the 
initial or final state. In the same works it has been shown that 
in some cases it is possible to recover factorization from a 
practical point of view. 
Systematic efforts 
have been undertaken\cite{JiMaYuan,JQVY06} 
for rewriting factorization rules for processes where partonic 
distributions depend on moderate values of the transverse momentum and  
the gauge factor is assumed to play a role. 

Several models and studies 
have been published on T-odd 
distributions\cite{Yuan03,GambergGoldsteinOganessyan03,
BacchettaSchaeferYang04,LuMa04,Pobylitsa03,DalesioMurgia04,Burkardt04,
Drago05,GoekeMeissnerMetzSchlegel06,BoerVogelsang06,Entropy3}, 
together with phenomenological 
parameterizations of the Sivers 
function\cite{Torino05,VogelsangYuan05,CollinsGoeke05,BR06a} 
and of the Boer-Mulders-Tangerman function\cite{Boer99,BaroneLuMa05}. 
So, we may say that T-odd distributions are an accepted reality, 
although we have no data yet about their behavior at asymptotic energies. 

This creates the problem of the coexistence of factorization breaking 
and factorization, i.e. at which 
extent it is possible to reconcile the interpretation of observable effects  
in terms of factorization breaking interactions with 
the safety of the extremely simple and useful 
factorization/universality scheme for classifying phenomenology. 
Also, it raises questions\cite{BrodskyEtAl02,Collins02}  
on the probabilistic interpretation
of the measured distributions. 
Most of the above quoted models assume, more or less 
explicitly, that 
the factorization breaking effects are there but small 
enough not to overthrow the underlying parton model picture. 
The analysis of \cite{BBMP} 
and the observations by \cite{CollinsQiu07} 
suggest that this point is still to be clarified, 
especially in hadron-hadron single spin asymmetries. 

Without considering a model in detail, we show here that 
it is possible to restore an approximate factorization scheme 
involving 
the presence of nonzero T-odd distribution functions, 
if factorization breaking interactions satisfy certain 
restrictive conditions. 

The central object of any distribution function calculation 
is the imaginary part of an amplitude $G(2,1)$ that 
describes the formation of a quark/hole pair at a spacetime 
point 1, and the propagation of the hole up to the point 
2 where the hole is filled by the quark again. 
A value of the longitudinal fraction $x$ is associated with 
this quark hole. By definition $x(1)$ $=$ $x(2)$.  
In absence of factorization breaking interactions, 
this is true also in local sense: the quark hole conserves 
its initial longitudinal fraction along all the path from 1 to 2. 

Factorization breaking interactions spoil this property. 
In any precise model treatment of rescattering, 
one includes explicitly degrees of freedom that 
exchange $O(Q)$ amounts of momentum/energy with the quark 
hole (see e.g. the calculations in ref.\cite{JQVY06}), and 
consequently the wavefunction describing the quark hole 
contains plane wave components associated to values $x'$ 
$\neq$ $x$. A way to quantify the effectiveness of rescattering 
is to look at the size of the difference $x-x'$. 

Our starting point is the idea that a rescattering mechanism 
produces a finite but $small$ violation of $x-$conservation. 
To formalize this, in section IV we introduce 
the $x-$frequency spectrum 
$f_\epsilon(x-x')$ of the rescattering operator 
and require a set of properties for it. Qualitatively, we may  
write since now that we require $f_\epsilon(x-x')$ to  
consist of a regular and narrow peak at $x$ $=$ $x'$, 
with finite width $\epsilon$ $<<$ 1. 
In addition, we require $f_\epsilon(x-x')$ to respect 
causality (see section IV for details). 

The $\epsilon$ parameter is supposed to depend slowly 
on $x$ and on $|k_T|$ 
($\vec k_T$ is the quark transverse momentum), and it expresses 
the magnitude of $x-$nonconservation in 
rescattering: $|x-x'|$ $\sim$ $\epsilon$. 
In the limit where $\epsilon$ $\rightarrow$ 0, rescattering 
does not affect the quark longitudinal fraction, and factorization 
is fully respected. 

That the frequency spectrum is large for $x'$ $\approx$ $x$, and small 
when $x$ and $x'$ are very different, is suggested 
by the fact that at a certain extent factorization 
works empirically. 
The assumption that the peak of $f_\epsilon(x-x')$ at $x'$ $=$ $x$ 
is regular is possibly not justified for 
individual perturbative diagrams, but becomes more reasonable when 
we consider resummed sets of diagrams, 
or approximations where 
some degrees of freedom have been integrated over, 
or diagrams that are regularized 
via cutoffs, form factors, and so on. 
Clearly, the chosen values for model parameters may critically 
influence the size of the relevant region of the frequency 
spectrum. As above written, here we suppose that this 
size $\epsilon$ is small. 

If the rescattering operator frequency spectrum satisfies the 
requested properties,  
on the ground of simple arguments 
we show that 
it is possible to write a transverse momentum dependent 
quark distribution in the form 
$q_0$ $+$ $\epsilon q_1$ $+$ $O(\epsilon^2)$, where $q_0$ may be 
calculated 
in complete absence of factorization breaking terms, and $q_1$ 
is a T-odd correction. 

This restores approximate universality for the T-even term: for 
values of $\epsilon$ up to 0.3 it may be rather difficult to 
detect deviations from the $q_0$ behavior. This also agrees 
with the 
known complete absence of screening effects in hadron-nucleus  
Drell-Yan (see \cite{Kenyon82} for a review of the related experiments). 

On the 
contrary, the T-odd correction term is potentially 
process-dependent unless one is able to demonstrate that 
$\epsilon$ has universal features. An effort in this 
direction has been undertaken 
elsewhere\cite{Entropy3} by the author of the present work, 
but the arguments presented here are 
independent from the analysis in \cite{Entropy3} and so we do not put 
constraints on $\epsilon$ in the following. 

As a last remark, the enlarged-delta approximation\cite{BR97} 
used in the following was developed to quickly approximate interactions 
between a pointlike particle and a composite system coexisting at 
the same time. 
Although it is probably possible to 
re-examine it, and to extend its validity to the case where the interacting 
particles exist at different times (the case of lepton-hadron SIDIS, 
and interactions between initial and final hadronic systems in 
hadron-hadron SIDIS), the present analysis is 
limited to the case of Drell-Yan and hadron-hadron initial state 
interactions. 

\section{Some general definitions and notations} 

We work, as far as possible, in the traditional 
picture of a distribution 
function, where a quark/hole pair is formed in a proton state 
$\vert P>$ at a spacetime point ``1'' $\equiv$ $(0,0,0,0)$,  
and the hole propagates up to a point ``2'' $\equiv$
$(0,z_-,\vec b)$, where 
the quark is restored with the same momentum $k_\mu$ 
it had in the origin. 
The imaginary part of this amplitude gives, by suitable Dirac 
projections, 
all the distribution functions we need. 

The quark/hole longitudinal fraction $x$ is defined by $P_+$ $=$ $x k_+$, 
where $P_+$ and $k_+$ are the large light-cone momentum components 
of the parent proton and of the quark. 
The infinite momentum limit $P_+$ $\rightarrow$ $\infty$ selects 
leading twist contributions. 
Not to work with 
singularities, we introduce the scaled longitudinal variable $\xi$: 

\noindent
\begin{equation}
\xi\ \equiv\ P_+ z_-, \label{eq:def_xi}
\end{equation}

and a distribution function may be written as 

\noindent
\begin{equation}
q(x,\vec k_T)\ \equiv\ P_+ \int e^{-i xP_+z_-} e^{ik_T b}
g(z_-,\vec k_T) dz_- d \vec k_T\ \equiv\ 
\int e^{-i x\xi} e^{ik_T b} 
G(\xi,\vec b) d\xi d \vec k_T. \label{eq:def_qx}
\end{equation}

The range of useful values of $\xi$ remains 
finite $\sim$ $1/x$ when $P_+$ $\rightarrow$ $\infty$. 
So in this limit 
the function $G(\xi,\vec b)$ does 
not need to be singular in the origin to produce a nonzero $q(x)$.  

The coordinate $z_+$ plays no role in the following and 
is not explicitly reported anymore. 
%We name $A(+)$ and $A(-)$ those areas of the process/diagram 
%associated with (hypothetically) factorization-separated particles. 
%The area of our interest is $A(+)$, where the relevant 
%components are $+$ (momentum), and $-$ (spacetime). 
%We will ignore completely what happens inside $A(-)$. 
%We will however assume that 4-momentum exchanges between $A(+)$ 
%and $A(-)$ take place, undermining factorization. 

%As a last point, factors like $\sqrt{2}$ or 
%$\pi$ will be normally ignored. 

\section{Enlarged $\delta$ function}

We adapt here a technique from refs. \cite{BR97} and \cite{BR_JPG} 
to roughly but quickly approximate the resummed effect of rescattering.
Rescattering obliges us 
to include nonlocality with respect to $x$, involving 
integrals of the kind $\int dx' f(x,x') h(x')$ where 
the kernel $f(x,x')$ becomes $\delta(x-x')$ in the limit of 
respected factorization. In this work we consider kernels that 
allow for a $small$ $x-$nonlocality: $\vert x-x'\vert$ $\sim$ 
$\epsilon$ $<<$ 1. 

Given a small nonzero parameter $\epsilon$ $<<$ 1,  
we define an ``$\epsilon$-enlarged $\delta-$function'' 
$\delta_\epsilon(x-x')$ such as to satisfy: 

(i) $x'$ $\approx$ $x$ within uncertainty 
$\vert \epsilon \vert$, 

(ii) $\int_{-\infty}^\infty dx' \delta_\epsilon(x-x')$ $=$ $1$.  

\noindent
To define $\delta_\epsilon$ 
in the limited $x-$range $[0-1]$, (ii) is better modified 
into: 

(ii b) $1-\alpha\epsilon^2$ $<$ 
$\int_0^1 dx' \delta_\epsilon(x-x')$ 
$<$ 1, with $\alpha$ $\sim$ 1. This property is valid only 
for $x$ $>$ $\epsilon$ and $x$ $<$ $1-\epsilon$, so values 
of $x$ near 0 and 1 will not be considered in this work. 

Such functions are 
normally used to define the ordinary $\delta(x-x')$ 
function in the 
$\epsilon$ $\rightarrow$ 0 limit. 
Several functional choices for $\delta_\epsilon(x-x')$ 
are possible, the most useful for us is (for $\epsilon$ $>$ 0, else 
$\epsilon$ $\rightarrow$ $\vert \epsilon\vert$):  

\noindent
\begin{equation}
\delta_{\epsilon}
(x-x')\ \equiv\ 
{1\over \pi} {\epsilon \over {\epsilon^2 + (x-x')^2}} 
\nonumber
\end{equation}
\begin{equation}
\equiv\ \delta_{\epsilon+}(x-x')\ 
+\ \delta_{\epsilon-}(x-x')
\equiv\ 
{i\over {2\pi}} {1 \over {x-x' + i\epsilon}}\ -\ 
{i\over {2\pi}} {1 \over {x-x' - i\epsilon}}. 
\end{equation} 

Causality implies the substitution 
\noindent
\begin{equation}
\delta_{\epsilon}(x-x')\ \rightarrow 
\delta_{\epsilon\pm}(x-x'). 
\end{equation}
that corresponds, in its Fourier transform, to 
\begin{equation}
exp(-\epsilon \vert \xi \vert)\ \rightarrow 
\theta(\pm\xi)exp(\mp\epsilon \xi) 
\end{equation} 

Since all the relations interesting us contain a function 
$\theta(\xi)$ related with causality, 
$\delta_\epsilon$ actually means $\delta_{\epsilon+}$, 
i.e. one pole only is encircled in the complex plane integrations. 

We are especially interested in the calculation of integrals 
of the form $\int_0^1 dx' f(x') \delta_{\epsilon+}(x-x')$. 
For $\epsilon$ $<<$ 1, and $x$ far from its kinematic  
limits 0 and 1 ($x$ $>$ $\epsilon$ and $1-x$ $>$ $\epsilon$)
such integrals  
may be assumed to be dominated by the $\delta_{\epsilon+}$ pole  
$x'$ $=$ $x+i\epsilon$. 

More precisely, we map 
the $[0,1]$ $x-$range onto the $[-\infty,+\infty]$ $z-$range 
by the transformation 

\noindent
\begin{equation}
x\ \equiv\ {e^z \over {1+e^z}}\ \leftrightarrow\ z\ \equiv\ ln(x/1-x)
\label{eq:map1}
\end{equation}

\noindent
Since $dx$ $=$ $e^z/(1+e^z)^2dz$, and 
for $x$ near the complex point $a$ we have  
$1/(x-a)$ $\approx$ 
$(1+e^z)^2/e^z\cdot 1/(z-A)$ 
(where $A$ $\equiv$ $z(a)$) 
we have 

\noindent
\begin{equation}
{{dx} \over {x-a}}\ \approx\ {{dz} \over {z-A}}  
\hspace{0.5truecm}for \ x\ \approx\ a. 
\label{eq:map2}
\end{equation}

If the integral 
$\int_0^1 f(x') \delta_{\epsilon+}(x-x')$ is dominated by the 
$x'$ $=$ $x'-i\epsilon$ singularity we have: 

\noindent
\begin{eqnarray}
\int_0^1 f(x') \delta_{\epsilon+}(x-x') dx'\ =\  
{1 \over {2\pi i}} 
\int_{-\infty}^\infty {{f[x'(z')]} \over {[z' - z(x+i\epsilon)]}} dz'
\\ 
\ =\ f(x + i\epsilon) 
\ =\ f(x)\ + {i \epsilon} {{df} \over {dx}}\ +\ O(\epsilon^2). 
\label{eq:approx1}
\end{eqnarray}

\section{Insertion of factorization breaking interactions} 

Let us define 

\noindent
\begin{equation}
q_0(x)\ \equiv\ 
\int d\xi e^{-ix\xi} 
< P \vert \psi(\xi)\psi^+(0) \vert P >,   
\label{eq:q_0} 
\end{equation}

\noindent
\begin{equation}
q_\epsilon(x)\ \equiv\ 
\int d\xi e^{-ix\xi} 
< P \vert \psi(\xi) F_\epsilon(\xi) \psi^+(0) \vert P >.   
\label{eq:q_epsilon} 
\end{equation}

Of the previous two equations, the former defines a parton-model 
leading twist quark distribution, the latter includes the effect of 
a rescattering operator $F_\epsilon(\xi)$, where a parameter 
$\epsilon$ quantifying the strength of $F$ is explicitly 
reported. For $\epsilon$ $=$ 0, 
$F_\epsilon$ reduces to unity, and $q_\epsilon$ to the parton 
model definition. 

We assume that $F_\epsilon$ is scalar, only to simplify the following 
relations (else, a double sum is present, see the end of this 
section). We introduce the frequency spectrum $f_\epsilon(x-x')$: 

\noindent
\begin{equation}
F_\epsilon(\xi)\ \equiv\ 
\int dy e^{-iy\xi} f_\epsilon(y), \hspace{0.5truecm} -1\ <\ y\ <\ 1. 
\label{eq:freq_spectr0} 
\end{equation}

\noindent

For $f_\epsilon(y)$ 
we assume the following properties: 

1) Although $f_\epsilon(y)$ is not a delta function, 
it consists of a peak at $y$ $=$ 0. 

2) This peak is regular on the real axis. 

3) The width $\epsilon$ of the peak is small: 
$\epsilon$ $<<$ 1. 

4) $f_\epsilon(y)$ is negligible for $|y|$ $>>$ $\epsilon$. 

5) For $\epsilon$ $\rightarrow$ 0, $f_\epsilon(y)$ $\rightarrow$ 
$\delta(y)$. 

6) The shape of this frequency spectrum is causality-modified, 
so that $F(\xi)$ contains the factor $\theta(\xi)$. In other words, 
$f_\epsilon(y)$ is not an ordinary function but a distribution, 
associated with given integration rules. 

The previous hypotheses allow one to approximate $f_\epsilon(y)$ 
the following way: 

\noindent
\begin{equation}
f_\epsilon(y)\ \approx\ 
\delta_{\epsilon+}(y). 
\label{eq:freq_spectr} 
\end{equation}

Now we may write 

\noindent
\begin{eqnarray}
q_\epsilon(x)\ =\ 
\int d\xi e^{-ix\xi} 
< P \vert \psi(\xi) 
\int dy e^{-iy\xi} \delta_{\epsilon+}(y) 
 \psi^+(0) \vert P > 
\ =\\
=\ 
\int dy \delta_{\epsilon+}(y) \int d\xi 
e^{-i(x+y)\xi} 
< P \vert \psi(\xi) \psi^+(0) \vert P > 
\ =\\
=\ \int dy \delta_{\epsilon+}(y) q_0(x+y)
\ =\ 
\int dy \delta_{\epsilon+}(x-y) q_0(y).
\label{eq:dist_q1} 
\end{eqnarray}

Applying the approximation eq.\ref{eq:approx1} we get 

\noindent
\begin{equation}
q_\epsilon(x)\ \approx\ q_0(x + i\epsilon)\ 
\approx\ q_0(x)\ 
+\ {i \epsilon} {{dq_0} \over {dx}}\ +\ O(\epsilon^2). \label{eq:dist_q2} 
\end{equation}

If one wants to be more precise and consider the non-scalar 
operator nature of $F_\epsilon$, one may introduce explicitly 
the matrix sum 
$<P\vert \sum_{i,j}....\vert P>$, 
and apply the previous 
considerations to each of the terms of the sum. 

\section{Detection of the finite imaginary part of the amplitude pole 
as a T-odd distribution} 

For making the above imaginary shift observable in the form of 
a T-odd $\vec k_T-$dependent distribution function,  
factorization breaking interactions 
must also produce nonzero interference between orbital states differing 
by one unit. 

So, we apply the considerations of the previous subsection to 
the impact parameter dependent distribution $q(x,\vec b)$. 

As a reference example, we consider the Sivers distribution. 
A Sivers contribution in agreement with the Trento definition\cite{Trento}
may be introduced in the unpolarized quark distribution of a 
transversely polarized proton as 

\noindent
\begin{equation}
q(x,k_x,0) \ =\ 
q_U(x,k^2)\ +\ {k_x \over M} q_S(x,k^2),
\label{eq:trento_sivers} 
\end{equation}

\noindent
where we have assumed that the proton is fully polarized along the 
$\hat y$ axis. 

We may Fourier-transform this equation with respect 
to $\vec k$, for $k_y$ $=$ 0, and write the result in the form  

\noindent
\begin{equation}
q(x,\vec k)\ =\ 
\int d^2 b e^{i b_x k_x} 
\Big(
Q_U(x,b^2)\ +\ i b_x {\partial \over {\partial (b^2)}} Q_S(x,b^2)
\Big)
\ \equiv\ 
\int d^2 b e^{i b_x k_x} 
\Big(
Q_U(x,b^2)\ +\ i b_x \epsilon''(b^2) Q_S''(x,b^2)
\Big)
\label{eq:Qb} 
\end{equation}

\noindent
where $Q_{U,S}$ is the Fourier transform of $q_{U,S}$ with respect to 
$exp(i\vec b \cdot \vec k)$, and all Q-functions are real and 
$\vec b-$even. 

If we apply the above eq.\ref{eq:dist_q2} to the impact parameter 
dependent distribution, and explicitly separate the $b_x-$even and 
$b_x-$odd parts of $\epsilon$: 

\noindent
\begin{equation}
\epsilon\ \equiv\ \epsilon'(b^2)\ +\ b_x\epsilon''(b^2), 
\label{eq:epsilon1} 
\end{equation}

\noindent
\begin{eqnarray}
Q_\epsilon(x, \vec b)
\ \approx\ 
Q_0(x + i\epsilon, b^2)\ 
\approx\ Q_0(x, \vec b)\ 
+\ {i \epsilon} {{dQ_0(x,b^2)} \over {dx}}
\\ \equiv\ 
Q_0(x, b^2)\ 
+\ {i \epsilon'(b^2)} 
{{d Q_0(x,b^2)} \over {dx}}
+\ {i b_x \epsilon''(b^2)} 
{{d Q_0(x,b^2)} \over {dx}},
\label{eq:distq3} 
\end{eqnarray}

\noindent 
it is evident that a part of the frequency spectrum of the 
interaction operator contributes to the 
Sivers term in eq.\ref{eq:trento_sivers}. 

\section{Conclusions} 

In this work, we have considered the action of a set of 
factorization breaking interactions, whose frequency spectrum 
is regular and peaked near the condition of conserved $x$. 
We required the width of this peak to be small: 
$\delta x$ $\approx$ $\epsilon$ $<<$ 1. 

We have shown that for such interactions it is possible 
to retain the factorization scheme, and 
write the result of the calculation of a distribution function 
as a series expansion with 
respect to $\epsilon$. 

The zeroth-order term is the T-even and universal distribution 
function $q(x)$ corresponding to the  
complete absence of initial state 
interactions. T-even corrections to $q(x)$ appear 
at second order in $\epsilon$. 

The first order correction is a T-odd distribution 
function, detectable in angular even-odd interference terms. 
It cannot be stated to be universal, at this level of analysis. 

%%%%%%%%%%%%%%%%%%%%%%%%%%%%%%%%%%%%%%%%%%%%%%%%%%%%%%%%%%%%%%%%%%%%%%%%%%%%

%%%%%%%%%%%%%%%%%%%%%%%%%%%%%%%%%%%%%%%%%%%%%%%%%%%%%%%%%%%%%%%%%%%%%%%%%%%%%%

\end{document}